\documentstyle[psfig]{mn}

\begin{document}
\title[z=2.15 Red Quasar Companions]{Red Companions to a z=2.15 Radio Loud Quasar}
\author[D.L. Clements]
{D.L. Clements$^{1,2,3}$\\ $^1$Department of Physics and Astronomy,
University of Wales Cardiff, PO Box 913, Cardiff, CF24 3YB\\
$^2$Institut d'Astrophysique Spatiale, B\^atiment 121, Universite
Paris XI, F-91405 ORSAY CEDEX, France\\ $^3$European Southern Observatory,
Karl-Schwarzschild-Strasse 2, D-85748 Garching-bei-Munchen, Germany}

\maketitle

\begin{abstract}
We have conducted observations of the environment around the z=2.15
radio loud quasar 1550-269 in search of distant galaxies associated
either with it or the z=2.09 CIV absorber along its line of
sight. Such objects will be distinguished by their red colours,
R-K$>$4.5. We find five such objects in a 1.5 arcmin$^2$ field around the
quasar, with typical K' magnitudes of $\sim$20.4 and no detected R band
emission. We also find a sixth object with K=19.6$\pm$0.3, and
undetected at R, just two arcseconds from the quasar. The nature of all these
objects is currently unclear, and will remain so until we have determined
their redshifts. We suggest that it is likely that they are associated with
either the quasar or the CIV absorber, in which case their properties
might be similar to those of the z=2.38 red Ly$\alpha$ emitting galaxies
discovered by Francis et al. (1997). The small separation between
the quasar and the brightest of our objects suggests that it may be the
galaxy responsible for the CIV metal line absorption system. The closeness
to the quasar and the red colour might have precluded similar objects from
being uncovered in previous searches for emission from CIV
and eg. damped absorbers.
\end{abstract}


\section{Introduction}

The selection of high redshift galaxies on the basis of their colours
has been a growth industry over the last 5 years. Much of this
work has concentrated on the selection of high (z$>$3) redshift
objects through `dropout' techniques. Such methods have had
considerable success (eg. Steidel et al. 1999). However, many of these
techniques are reliant on emission in the rest-frame ultraviolet. The
UV emission from a galaxy can easily be dominated by a small burst of
star formation, or alternatively obscured by a relatively small amount
of dust. A population of older quiescent galaxies might thus coexist
with the UV selected high redshift objects. Studies of the stellar
populations in moderate redshift radio galaxies provide some support
for this idea.  A number of authors (eg. Stockton et al. (1995),
Spinrad et al. (1997)) have shown that several radio galaxies have
ages $>$3--5 Gyr at z$\sim$1.5, indicating that they must have formed
at z$>$5. These results have even been used (Dunlop et al. 1996) to
argue that $\Omega$ must be significantly less than 1.

Old galaxies at moderate redshift, passively evolving from z$>$5 to
z=1.5 -- 2.5, would appear as red objects, with R-K' colours
$>$4.5. There has been considerable interest in such red objects. Much
of this work has centred on red objects found in the fields of known
high redshift AGN (eg. Hu \& Ridgeway (1994), Yamada et al.,
(1997)). A large survey of the environments of z=1--2 quasars (Hall et
al., 1998) finds that such associations are quite common. The present
paper attempts to push such studies above z=2. The alternative
approach, to study red objects in the field, is also an active area
with several surveys dedicated to or capable of finding such objects.
See eg. Cohen et al. (1999), or Rigopoulou et al. (in preparation).
Red objects need not be old, though. An alternative explanation is that they
are heavily obscured, and may contain either a redenned AGN or massive
starburst (eg. Dey et al., 1999, Egami et al., 1996). In this context it is
interesting to note that several of the objects found in recent
deep submm surveys have been identified with very red objects (Smail
et al., 1999; Clements et al., in preparation).

Finding emission from the putative galaxies responsible for metal and
damped-Ly$\alpha$ absorption line systems has been the goal of
numerous observational programmes. At low redshift there has been
considerable success in identifying the galaxies responsible for MgII
absorption systems (Bergeron \& Boisse, 1991; Steidel et al., 
1997. At higher redshifts, interest has mostly focussed on the
damped-Ly$\alpha$ absorption systems. Searches for line emission from
such objects (eg. Bunker et al. 1999; Wolfe et al. 1992) has met with
varying success (Leibundgut \& Robertson, 1999). Fewer observers have looked in the
continuum, but there have been some successes there as well. For
example, Aragon-Salamanca et al. (1996) found close companions to 2
out of 10 quasars with damped absorbers in a K band survey. As yet
there has been no spectroscopic confirmation of these identifications,
but the broad characteristics of these objects, and the small fraction
of damped absorbers detected, is consistent with plausible models for
the evolution of the galaxies responsible (Mathlin et al., in
preparation). Meanwhile, Aragon-Salamanca et al. (1994) looked for
counterparts to multiple CIV absorbers lying at z$\sim$1.6, also using K band
observations. They found an excess of K band objects near to the quasars,
consistent with their being responsible for the CIV absorption. Once again,
there is no spectroscopic confirmation of the assumed redshifts.

The present paper presents the first results of a programme aimed at
finding quiescent objects at high redshift (z$\sim$2--2.5) using
optical/IR colour selection techniques. Among the targets observed in
an initial test programme was the radio loud quasar 1550-2655,
selected as an example radio loud object. The rest of the paper is
organised as follows. Section 2 describes our observations, data
analysis and presents the results. Section 3 discusses these results
and examines three possible origins for the red objects we have found
to be associated with 1550-2655. Finally we draw our conclusions.  We
assume $\Omega_M$ = 1, $\Lambda$=0 and H$_0$=100 kms$^{-1}$Mpc$^{-1}$
throughout this paper.

\section{Observations and Results}

As part of a programme to examine the role of quiescent galaxies at
z=2--2.5, we observed the field surrounding the radio loud quasar
1550-2655. This object lies at a redshift of 2.15 and shows signs of
associated Ly$\alpha$ absorption (Jauncey et al., 1984). Its spectrum
also contains a CIV absorber at z=2.09. Observations were made at
the 3.5m ESO NTT, and data reduction used standard IRAF and Eclipse
routines.

The optical observations, in R band, were conducted in service mode on
20 August 1997 using the SUSI imager. This provides high resolution
images, with a pixel size of 0.13''.  A total integration time of
3600s was obtained on the source. This integration time was broken up
into 12 subintegrations of 300s each, whose relative positions were
shifted by up to 40'' in a semi-random jitter pattern. These images
were bias subtracted, flat fielded using a sky flat made on the
twilight sky, then aligned and median combined to produce the final
image.  A residual gradient going from left to right across the image
was apparent in the final data. This was removed by subtracting a
low-order polynomial fitted to each horizontal row of pixels once
detected objects had been masked off. The final R band image is shown
in Figure 1a. Seeing was measured to be marginally subarcsecond on the
final image.

The infrared data was obtained using the K' filter on the SOFI
infrared imager on 12 July 1998. The 3600s of integration were obtained
in 60 one minute sub-integrations which themselves were the result of
six 10s integrations.  The 60 one minute sub-integrations were shifted
relative to one another in a random 15'' size dithering pattern to
allow for sky background determination and subtraction. The flat field
was obtained using a standard lamp ON $-$ lamp OFF dome flat. The
data were reduced using the Eclipse package by Nick Devillard
(1997). The algorithms used for reducing dithered infrared data in
this package are detailed in Devillard (1999). In summary, the package allows
for flat fielding with the preprepared flat, and conducts sky
subtraction using a running average of 10 offset images. It then
identifies sources on each image and uses a correlation technique to
calculate the offsets between them. The separate subintegrations are
then offset and combined to produce the final image. Seeing was measured in
the final image to be $\sim$ 0.9''.

Photometric calibration used the Landolt standard
(Landolt, 1992) PG1633+099C in R band, and the faint IR standards
P499E and S875C at K' (Cassali \& Hawarden, 1992). Galactic extinction
was corrected using values obtained from the NASA Extragalactic
Database (NED).

After data reduction and flux calibration, the SUSI image, which has a
resolution of 0.13''/pixel, was rebinned to match the 0.292''/pixel
SOFI resolution, and the images were aligned. The final matched images
were 67 by 88 arcseconds in size. The main limiting factor on this
size was the small SUSI field of view and the dithering scheme used
for the optical observations. We then used SExtractor to select
objects detected at K' and to extract their photometric properties in
matched apertures in the two passbands. To qualify for detection, an
object had to have a 1.5$\sigma$ significance flux in 10 connected
pixels in the K' band image (ie. $\sim$ 5$\sigma$ significance
overall). This matched catalogue can then easily be searched for
objects with specific colour criteria.  We detected a total of 75
objects in K' down to a limiting magnitude of $\sim20.5$.

The catalogue was then searched for candidate red objects, with
R-K'$>$4.5.  We found five such objects in the catalogue, details of
which are given in Table 1. Their positions are also shown in Figure 1,
which shows both the R and K' band images of the quasar field.

\subsection{A Red Quasar Companion}

Comparison of the R and K' images of the quasar itself shows what would
appear to be a red companion object --- apparent as an extension in K,
but absent in the R band image. The reality of this object was
investigated by subtracting off the unresolved quasar contribution.
This was achieved by selecting a star, with no close companions, in
the observed field and using this as a PSF model. The central value of
the PSF image was scaled to match that in the quasar image, and then
the two images were subtracted. The companion was clearly visible in
the K'-band PSF subtracted image, but was entirely absent in the R-band
PSF subtracted image. The companion is marginally resolved, having a
size of roughly 1.5 x 1 arcseconds, and is situated $\sim$2 arcseconds
from the quasar.  R and K' magnitudes were extracted from the PSF
subtracted image, indicating that the quasar companion is also
red. Its details are included in Table 1.

\vspace{1cm}
\begin{table}
\begin{center}
\begin{tabular}{crrr} \hline
Object No.&Kmag&Rmag&R-K\\ \hline
R1&20.2$\pm$0.1&$>$24.8&$>$4.6\\
R2&20.2$\pm$0.1&$>$25.3&$>$5.1\\
R3&20.4$\pm$0.1&$>$25.3&$>$4.9\\
R4&20.3$\pm$0.1&$>$24.9&$>$4.6\\
R5&20.3$\pm$0.1&$>$25.0&$>$4.7\\
C1&19.5$\pm$0.3&$>$24.5&$>$5.0\\ \hline
\end{tabular}
\caption{Properties of red companions to the RLQ 1550-2655}
All limits given are 3$\sigma$. The quasar companion is object C1.
\end{center}
\end{table}

\begin{figure*}
\begin{tabular}{cc}
\psfig{file=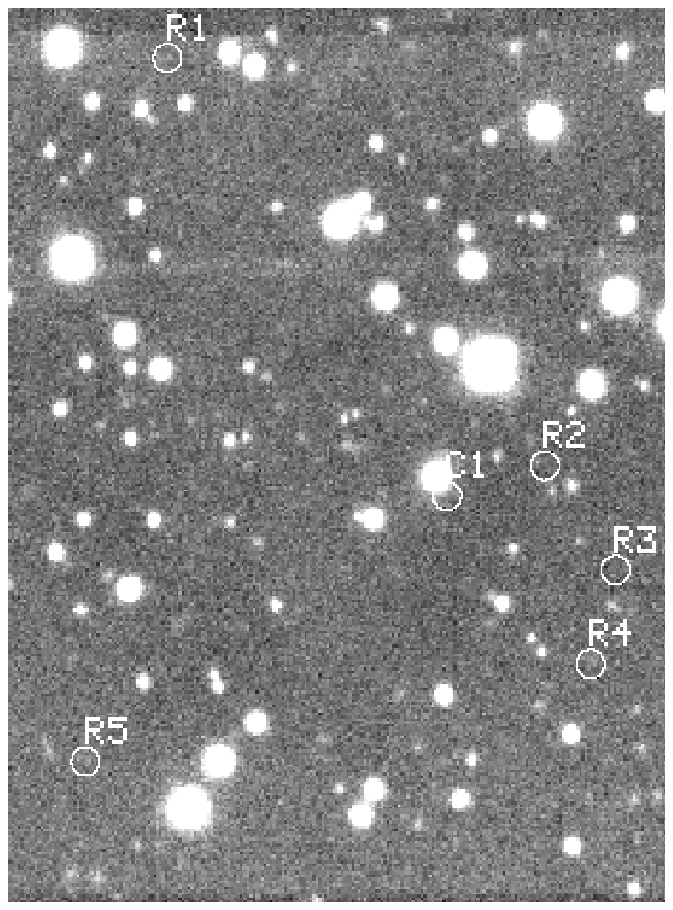,width=8cm}
\psfig{file=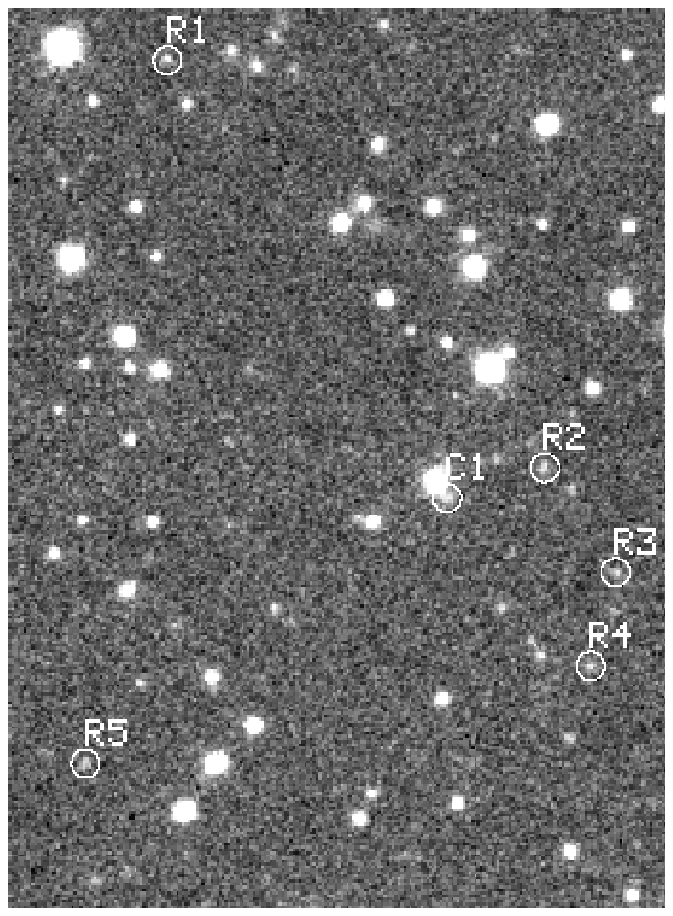,width=8cm}\\
\end{tabular}
\caption{R (left) and K' band (right) images of the quasar field.}
The red associates are indicated in each image. Note their clear detections
in the K' image, and their non-detections in the deeper R band image.
\end{figure*}

\section{Discussion}

Until we can obtain spectroscopy for these objects, it is difficult to
assess their importance or role in this system. There are three possible
origins for these red objects: (1) they are associated with the quasar,
lying at z=2.15; (2) they are associated with the CIV absorber at
z=2.09; (3) they are foreground objects, unrelated to the quasar or CIV
system. We will assess each of these alternatives in turn.

\subsection{Association with the Quasar}

The density of red objects in this field is surprisingly high -
$\sim 4\pm1.5$ arcmin$^{-2}$ as opposed to the supposed global value of
$\sim 1\pm0.3$ arcmin$^{-2}$ (Cohen et al. 1999). The density of red objects
near to the quasar is significantly higher, with three of the red
objects as well as the companion lying in a 0.25 arcmin$^{-2}$ region
near the quasar. This is certainly suggestive that there is some
connection between the red objects and the AGN (or CIV system).  Other
studies have found a similar connection between red objects and AGN,
especially radio loud AGN. Perhaps the largest survey to date is that
of Hall et al. (1999) who obtained images of 31 z=1 -- 2 radio loud
quasars and found a significant excess of red galaxies around them.
Interestingly they found two radial dependencies for this excess, one
of which lies close to the quasar ($<$40'') and another more distant
(40''--100''). This is perhaps reflected in the present study, with
two red objects in the second class, further from the quasar, and the
rest, including the close companion, within 40''. In this
interpretation, the close companion would be $\sim$26kpc from the
quasar, and would be about 20x15 kpc in size. Hydrogen in this object
might be responsible for the associated absorption seen in the quasar
spectrum.

If we take the redshift of the companion galaxies to be the same as
the quasar, then the implied absolute magnitudes would be $\sim$-24 to
-25, ie. about L$^{*}$ (using the luminosity function of Mobasher et
al., 1993 and converting to the assumed cosmology). This calculation
of course ignores K-corrections, but these are expected to be quite
low in the K-band. Cowie et al (1994) calculate K-corrections at
z$\sim$2.1 to be less than 1 for all morphological classes from E to
Irr.  We also note that these results are not dissimilar to those of
Francis et al. (1997) who uncovered a group of similarly red objects
at z=2.38 associated with a cluster of quasar absorption line
systems. K' magnitudes for these objects are similar to, or brighter
then, those of the objects discussed here.  It is interesting to note
that the Francis et al. objects are all Ly$\alpha$ emitters. If the
present red objects have similar properties, then such line emission
would make redshift determination much easier.

\subsection{Association with the CIV Absorber}

Many of the same comments regarding direct association with the quasar
can be made regarding association with the CIV absorber: there is an
unusually high density of red objects in this region suggesting some
connection between them. Of particular interest here is the closeness
of the quasar companion to the quasar - only $\sim$2'' away, or 26kpc
at the redshift of the CIV absorber, and with a size similar to that 7 

given above. To date little is known about the nature of CIV
absorbers, so the possible identification of the galaxy responsible
for one is rather interesting. Previous work looking for emission
lines from objects associated with damped and CIV absorbers (Mannucci
et al., 1998) suggests that galaxies are more likely to cluster with
absorbers than with quasars. If correct, this would suggest that the
objects found here are more likely to be associated with the absorber
than the quasar.

There has been a long and largely unsuccessful history of searches for
emission from absorption line systems in quasars at large
redshift. These have largely concentrated on emission lines, whether
Ly$\alpha$ (eg. Leibundgut \& Robertson, 1999), H$\alpha$, (eg. Bunker
et al., 1999), or others, though work in the infrared continuum has perhaps
shown greater success (see eg. Aragon-Salamanca et al., 1996,
1994). If the quasar companion in the present study is indeed
responsible for the CIV absorption, then we may have an explanation
for the failures.  The object is both red and quite close to the
quasar.  Detection of the companion would require both good seeing
(conditions for our own observations were sub-arcsecond) and
observations in the near-IR as well as the ability to subtract off the
quasar contribution. Sensitive infrared detectors have only recently
become available at most observatories, while subarcsecond seeing is
only rarely achieved. We might thus have been lucky in being able to
detect the companion. New instruments, such as UFTI at UKIRT, which
combines adaptive optics correction (regularly 0.5'') with a superb
infrared imager, can regularly make such observations. This will
hopefully allow us to make significant advances in our understanding
of high redshift absorption line systems.

\subsection{Foreground Contaminants}

The possibility that the red objects are at an entirely different
redshift to the quasar and absorber must still be considered while we
do not have confirming redshift spectra. In this context it is
salutary to note the lesson of the first VRO discovered (Hu \&
Ridgeway, 1994), known as HR10. This was found in the field of a
z=3.79 quasar, but was later shown to have a redshift of 1.44 (Graham
\& Dey, 1996).
However, the number density of red objects in their field was 0.9
arcmin$^{-2}$ which matches the field density of red objects discussed
by Cohen et al. (1999), and is lower than that found here.

\section{Conclusions}

At present there are several deficiencies in our data. Firstly we have
only obtained limits on the objects R band magnitudes. We must detect
them and measure, rather than limit, their R band magnitudes before we
can properly determine their colours. Secondly we must obtain spectra
for the objects so that we can actually determine, rather than
speculate on, their redshift. However, the results presented here
suggest that a larger survey of quasar environments, both with and
without absorbers, using infrared imagers with adaptive optics
correction might shed new light on galaxy populations at large redshift.
\\~\\
{\bf Acknowledgments} This paper is based on observations made at the
European Southern Observatory, Chile. It is a pleasure to thank Nick
Devillard for his excellent Eclipse data reduction pipeline, and
E. Bertin for SExtractor. This research has made use of the NASA/IPAC
Extragalactic Database (NED) which is operated by the Jet Propulsion
Laboratory, California Institute of Technology, under contract with
the National Aeronautics and Space Administration. I would like to
thank Amanda Baker and Garry Mathlin for useful discussions, and the
anonymous referee for helpful comments on an earlier version. This
work was supported in part by an ESO fellowship, EU TMR Network
programme FMRX-CT96-0068 and by a PPARC postdoctoral grant.

\end{document}